\documentstyle[aps,prb,epsf,multicol]{revtex}
 \tighten
   \begin{document}
   \title{\Large \bf{
         Floquet scattering theory of quantum pumps
                    }
         } 
\author{M. Moskalets$^{1}$ and
M. B\"uttiker$^2$
}
\address{
         $^1$Department of Metal and Semiconductor Physics,\\
        National Technical University "Kharkov Polytechnical Institute",
        Kharkov, Ukraine\\
        $^2$D\'epartement de Physique Th\'eorique, Universit\'e de Gen\`eve,
        CH-1211 Gen\`eve 4, Switzerland\\}

\date\today
   \maketitle 
   \bigskip   

   \begin{abstract}

We develop the Floquet scattering theory for quantum mechanical pumping 
in mesoscopic conductors. 
The nonequilibrium distribution function, the dc charge and heat currents
are investigated at arbitrary pumping amplitude and frequency.
For mesoscopic samples with discrete spectrum 
we predict a sign reversal of the pumped current 
when the pump frequency is equal to the level spacing in the sample.
This effect allows to measure the phase 
of the transmission coefficient through the mesoscopic sample.
We discuss the necessary symmetry conditions (both spatial and temporal)
for pumping.

   \end{abstract}

 \begin{multicols}{2}
 \narrowtext

\section{Introduction}
\label{i}
\indent

Quantum charge pumping 
\cite{SMCG99,Brouwer98,BTP94,HN91,ZSA99,SAA00,Simon00,WWG00,Brouwer01,PB01,MB01,AEGS00,AEGS01,Alekseev,VAA01,MM01,Levitov01,WWG02,CB02,EWAL02,blau,BWJW,MB02,PVB02,WW02,Kim02,Cohen}
is presently of considerable interest. 
An experiment by Switkes et al. \onlinecite{SMCG99} demonstrated that 
a phase coherent mesoscopic system subjected to 
a cyclic two parameter perturbation can produce a directed current.

Coherent quantum pumping is a consequence of the interference
of energetically different traversal paths 
made possible by an oscillating scatterer. 
The ratio of the oscillation frequency $\omega$ 
to the inverse time taken for carriers to traverse the sample 
$\tau_{T}^{-1}$ defines the operational regime of a pump \cite{note1,BL82,muga}.
Brouwer \cite{Brouwer98} gave
an elegant formulation of adiabatic 
($\omega\ll\tau_{T}^{-1}$) 
quantum pumping 
that is based on the scattering matrix approach to low frequency ac 
transport in phase coherent mesoscopic systems developed by
B\"{u}ttiker, Thomas, and Pr\^{e}tre \cite{BTP94}. 
This approach leads naturally to a geometrical description 
of adiabatic quantum pumping 
\cite{Brouwer98,AEGS00,AEGS01,Alekseev,VAA01,MM01,Levitov01}.
The theory predicts that 
the charge pumped during a cycle depends on the area enclosed 
by the path in the scattering matrix parameter space.
A less formal but more physical picture of 
an adiabatic quantum pump 
appeals to both quantum mechanical interference  
and photon assisted transport.
\cite{WWG02,MB02}.
The same processes are important for a nonadiabatic 
($\omega\gg\tau_{T}^{-1}$) pump 
\cite{SW96,Wagner,MPMB,WWG02}.
These discussions emphasize the energetics 
of the carrier traversal process. 

It is the purpose of this work to develop a theory that 
permits the description of both adiabatic and nonadiabatic regimes
on the same footing and allows a simple physical interpretation.
To this end we extend the approach of Ref.\onlinecite{MB02} to
the case of large frequencies and large pumping amplitudes. 
We apply the Floquet scattering theory 
\cite{Wagner,MPMB,Shirley65,LR99,MR01}
which deals with 
the scattering matrix dependent on two energies (incident and outgoing).
This approach leads to expressions for the quantities of interest 
in terms of the side bands of particles exiting the pump. 
The side bands \cite{BL82} correspond to particles 
which have gained or lost one or several modulation quanta
$\hbar \omega$. 
This approach is complementary to discussions based on
the scattering matrix dependent on two times 
\cite{VAA01,PVB02,WW02}

The paper is organized as follows.
In Sec.\ref{ga} the general approach to the kinetics of quantum pumps 
based on the Floquet scattering theory is presented.
In Sec.\ref{aa} we apply the general results to the adiabatic case.
In Sec.\ref{ch} we calculate the Floquet scattering matrix for a 
particular model - an oscillating double barrier potential - 
and present the results of numerical calculations of the 
pumped charge and the heat currents in both 
adiabatic and nonadiabatic regimes.
We conclude in Sec.\ref{dc}.

\section{General approach}
\label{ga}
\indent

We consider scattering 
\cite{Buttiker90,Buttiker92,BTP94}
of an incoming flow of electrons
with energy $E$ at a scatterer that oscillates in time 
with frequency $\omega$.

During the interaction with the oscillating scatterer \cite{BL82}
electrons can gain or lose energy quanta $\hbar\omega$. 
Hence the outgoing state is characterized by the set of 
energies $E_n$, $n = 0, \pm 1, \pm 2, \dots$,
\begin{equation}
     E_n = E + n\hbar\omega.
\label{Eq1_1}
\end{equation}
\noindent 
This is a Floquet state.

According to the Floquet theorem 
the energy ladder Eq.(\ref{Eq1_1}) gives the full set
of possible energies for outgoing particles
(see e.g., \cite{Wagner,LR99}).
Thus to describe scattering due to an oscillating scatterer
we can use the Floquet scattering matrix $\hat S_{F}$. 
The matrix element $S_{F,\alpha\beta}(E_n,E)$ 
is the quantum mechanical amplitude for an electron
with energy $E$ entering the scatterer 
through lead $\beta$ to 
leave the scatterer through lead $\alpha$ having absorbed
($n > 0$) or emitted ($n < 0$) energy quanta $|n|\hbar\omega$. 
The Greek letters $\alpha$, $\beta$,  number the leads
connecting the sample to $N_{r}$ reservoirs.

We remark that the negative values $E_n < 0$ 
correspond to bound states near the oscillating scatterer. 
These states influence scattering into the propagating 
($E_n > 0$) states but they do not directly contribute to the current.

Current conservation implies that the submatrix $\hat S^{(p)}_{F}$
of the Floquet scattering matrix
(corresponding to propagating modes only) is a unitary matrix
\begin{equation}
 \hat S^{(p)\dagger}_{F}\hat S^{(p)}_{F} = 
 \hat S^{(p)}_{F}\hat S^{(p)\dagger}_{F} = 
\hat I.
\label{Eq1_2}
\end{equation}
In particular, if a current with flux $1$ and energy $E$ enters
the scatterer through lead $\beta$ 
then current conservation implies

\begin{equation}
 \sum_\alpha \sum_{E_n>0} |S_{F,\alpha\beta}(E_n,E)|^2 = 1.
\label{Eq1_3}
\end{equation}
Another useful condition follows from the fact that if
all incoming propagating ($E_n > 0$) channels are full
then each outgoing channel has also to be full:

\begin{equation}
 \sum_\beta \sum_{E_n>0} |S_{F,\alpha\beta}(E,E_{n})|^2 = 1.
\label{Eq1_4}
\end{equation}
Note that usually the Floquet energy $E$ is determined within the
interval $0\leq E < \hbar\omega$. 
However, for our problem, it is convenient not to reduce the discrete set  
of $E_{n}$ to this interval and to keep $E$ as the actual energy of incident 
(or outgoing) particles.

Because of Eq.(\ref{Eq1_2}) we can express 
the annihilation operator $\hat b$ for outgoing particles in the lead $\alpha$
in terms of annihilation operators $\hat a$ for incoming particles 
in leads $\beta = 1,2,\dots,N_{r}$
as follows \cite{Buttiker92,MB02}

\begin{equation}
 \hat b_{\alpha}(E) = 
\sum_\beta \sum_{E_n>0} S_{F,\alpha\beta}(E,E_{n}) \hat a_{\beta}(E_n).
\label{Eq1_5}
\end{equation}

\noindent
The operators $\hat a_{\alpha}(E)$ 
obey the following anticommutation relations 
$$
 [\hat a^{\dagger}_{\alpha}(E), \hat a_{\beta}(E')] = 
\delta_{\alpha\beta}\delta(E - E').
$$
Using Eqs.(\ref{Eq1_2}) and (\ref{Eq1_5}) we see that the operators
$\hat b_{\alpha}(E)$ obey the same relations. 

Note that above expressions correspond to single (transverse) channel leads
and spinless electrons. For the case of many-channel leads 
each lead index ($\alpha$, $\beta$, etc.) includes a transverse channel 
index and any repeating lead  index implies implicitly a summation over 
all the transverse channels in the lead. 
Similarly an electron spin can be taken into account.

Now we calculate the distribution function 
$f^{(out)}(E)_{\alpha}$ $=$ 
$<\hat b^{\dagger}_{\alpha}(E)\hat b_{\alpha}(E)>$ 
for electrons leaving the scatterer through the lead $\alpha$.
Here $<...>$ means quantum-statistical averaging.
Taking into account Eq.(\ref{Eq1_5}) we obtain

\begin{equation}
\begin{array}{c}
f^{(out)}_{\alpha}(E) = \sum\limits_\beta \sum\limits_{E_n>0} 
|S_{F,\alpha\beta}(E,E_{n})|^2 \\
\ \\
\times f^{(in)}_{\beta}(E_n).
\end{array}
\label{Eq1_6}
\end{equation}

\noindent
Here 
$f^{(in)}_{\beta}(E_n)$ $=$
$<\hat a^{\dagger}_{\beta}(E)\hat a_{\beta}(E)>$ is the distribution function
for electrons entering the scatterer through lead $\beta$.

\subsection{Directed charge currents}
\indent
\label{cc}

Using the distribution function 
$f^{(out)}_{\alpha}(E)$ for outgoing particles and 
$f^{(in)}_{\alpha}(E)$ for incoming ones we can find the directed  current
$I_{\alpha}$ in the lead $\alpha$ far from the scatterer  \cite{MB02}

\begin{equation}
 I_{\alpha} = \frac{e}{h} \int\limits_{0}^{\infty} dE 
\left\{f^{(out)}_{\alpha}(E) - f^{(in)}_{\alpha}(E)\right\}.
\label{Eq1_7}
\end{equation}

\noindent
The current directed from the scatterer towards the reservoir 
is positive by definition.
Substituting Eq.(\ref{Eq1_6}) into the above equation, 
using Eq.(\ref{Eq1_4}), and making the shift $E \to E - n\hbar\omega$
we find 

\begin{equation}
\begin{array}{r}
 I_{\alpha} = \frac{e}{h} \int\limits_{0}^{\infty} dE 
 \sum\limits_\beta \sum\limits_{E_n>0} |S_{F,\alpha\beta}(E_{n},E)|^2 \\
\ \\
\times \left( f^{(in)}_{\beta}(E) - f^{(in)}_{\alpha}(E_{n})
\right).
\end{array}
\label{Eq1_8}
\end{equation}

\noindent 
Here 
$\sum_{E_n>0}$ means a sum over those $n$ (positive and negative)
for which $E_n = E + n\hbar\omega > 0$.

Another useful representation for the directed current can be obtained
if we use Eq.(\ref{Eq1_3}) and make the shift 
$E \to E - n\hbar\omega$
in $f^{(out)}_{\alpha}(E)$ 
in Eq.(\ref{Eq1_7}). 
As a result we obtain 

\begin{equation}
\begin{array}{l}
 I_{\alpha} = \frac{e}{h} \int\limits_{0}^{\infty} dE 
 \sum\limits_{\beta\neq\alpha} \sum\limits_{E_n>0} \\
\ \\
\left\{
|S_{F,\alpha\beta}(E_{n},E)|^2 f^{(in)}_{\beta}(E) 
\right. \\
\ \\
\left.
- |S_{F,\beta\alpha}(E_{n},E)|^2 f^{(in)}_{\alpha}(E)
\right\}.
\end{array}
\label{Eq1_9}
\end{equation}

\noindent
>From this expression for the directed 
current it follows that only transmission $\alpha\neq\beta$
(not reflection $\alpha=\beta$) contributes to the current.
In addition Eq.(\ref{Eq1_9})
can help us to consider the effect of 
time reversal symmetry (TRS) on the pumped current.

On the one hand, the time reversal $t\to -t$ (TR) interchanges incoming
and outgoing channels
$$
\left[ S_{F,\alpha\beta}(E_{n},E)\right]^{(TR)} = 
S_{F,\beta\alpha}(E,E_{n}).
$$
Hence if the TRS is present then Eq.(\ref{Eq1_9}) reads

\begin{equation}
\begin{array}{l}
 I_{\alpha}^{(TRS)} = \frac{e}{h} \int\limits_{0}^{\infty} dE 
f_{0}(E)
 \sum\limits_{\beta\neq\alpha} \sum\limits_{E_n>0} \\
\ \\
\left(
|S_{F,\alpha\beta}(E_{n},E)|^2  
- |S_{F,\alpha\beta}(E,E_{n})|^2
\right).
\end{array}
\label{Eq1_10}
\end{equation}

\noindent
In the above equation 
(in accordance with the usual pump setup) 
we suppose that incoming electrons 
in all the channels are described by the same Fermi distribution function
with temperature $T$ and electrochemical potential
$\mu$:
$$
f^{(in)}_{\alpha}(E) = f_{0}(E) \equiv\frac{1}{1 + 
\exp\left(\frac{E - \mu}{k_B T} \right)}.
$$

Generally, the Floquet scattering matrix elements for transmission 
with incident energy $E$ to $E_n$ is not equal 
to the transmission from $E_n$ to $E$, 
$$
S_{F,\alpha\beta}(E_{n},E) \neq
S_{F,\alpha\beta}(E,E_{n}).
$$
>From this we can conclude that 
even a pump with TRS can generate a directed current. 
If these two scattering amplitudes are not equal, 
there exists the possibility of empty states deep 
below the Fermi surface \cite{deep1,deep2}. 
In this case interaction and inelastic effects \cite{MB01} 
can be expected to be especially important. 

However, if the scattering matrix is energy
independent on the scale of the order of $\hbar\omega$ 
$$
S_{F,\alpha\beta}(E_{n},E) \approx
S_{F,\alpha\beta}(E),
$$
then the scatterer with TRS can not produce a dc current.
The last circumstance is especially important for the adiabatic case 
$\omega\to 0$, since the adiabatic scattering matrix 
Eq.(\ref{Eq2_4}) satisfies the above condition.

On the other hand, in general, the nonstationary case is without TRS
and the oscillating scatterer can, in principle, generate a current.
In particular, if two parameters of the scatterer oscillate in time
with the same frequency $\omega$ but with phase lag $\Delta\varphi$
then the time reversal implies the reversal of the sign of $\Delta\varphi$.
Thus for a pump with $\Delta\varphi\neq 0$ the scattering problem
is without TRS (the time reversal symmetry is dynamically broken)
and such a scatterer can produce a dc current.
This effect was used for generating a dc current 
in adiabatic quantum pumps \cite{SMCG99}.

\subsection{Directed heat currents}
\indent
\label{hc}

By analogy with Eq.(\ref{Eq1_7}) we find the directed heat current 
$I_{E,\alpha}$ flowing in lead $\alpha$ away from the scatterer
\cite{GBJB96,Moskalets98,Krive99,MB02}
(we suppose that all the reservoirs are at the same electrochemical 
potential $\mu$)

\begin{equation}
\begin{array}{l}
 I_{E,\alpha} = \frac{1}{h} \int\limits_{0}^{\infty} dE (E-\mu) \\
\ \\
\times\left\{f^{(out)}_{\alpha}(E) - f^{(in)}_{\alpha}(E)\right\}.
\end{array}
\label{Eq1_11}
\end{equation}

\noindent
With the distribution function for outgoing particles
$f^{(out)}$ from Eq.(\ref{Eq1_6}) we get

\begin{equation}
\begin{array}{l}
 I_{E,\alpha} = \frac{1}{h} \int\limits_{0}^{\infty} dE 
 \sum\limits_\beta \sum\limits_{E_n>0} (E_n - \mu) \\
\ \\
\times|S_{F,\alpha\beta}(E_{n},E)|^2 
\left( f^{(in)}_{\beta}(E) - f^{(in)}_{\alpha}(E_{n})
\right).
\end{array}
\label{Eq1_12}
\end{equation}

\noindent
Note that if all the reservoirs are
at the same macroscopic conditions 
(electrochemical potential, temperature, etc.)
then the heat flow $I_{E,\alpha}$
(at any lead $\alpha = 1,2,\dots, N_{r}$) 
is directed from the scatterer to the reservoir \cite{MB02}.
That differs strongly from the charge current 
$I_{\alpha}$ given by Eq.(\ref{Eq1_8})
which, if it exists, can be directed 
either from the reservoir to the scatterer (at some lead) or vice versa
(at another lead).

\section{Adiabatic approximation}
\indent 
\label{aa}

In this section we use the above formalism to investigate 
the limit of adiabatic scattering. 

The general physical notion of adiabaticity applied 
to the scattering problem of interest here is as follows:  
Let us suppose that the time independent problem is described by 
the scattering matrix $\hat S_{0}(E,X_1,\dots,X_N)$ which 
depends on the energy $E$ of incident electrons and a set of 
parameters $X_i$, $i = 1,2,\dots,N$. 
Next assume that the parameters $X_i$ vary in time:
$X_i = X_i(t)$. 
Then we can say that the nonstationary scattering problem is {\it adiabatic} 
if scattering of particles incident with energy $E$  
can be described via the scattering matrix
$\hat S_0$ with time-dependent parameters $X_i(t)$

\begin{equation}
\hat S_{ad}(E,t) = \hat S_0(E,X_1(t),\dots,X_N(t)).
\label{Eq2_1}
\end{equation}

\noindent 
This approximation is adequate if the scattering matrix
changes only a little while an electron interacts with the scatterer.
In other words, the characteristic time scale for the change of parameters
$X_i$ is much larger than the traversal time $\tau_{T}$. 

An 
analogous criterion can be formulated concerning the energy dependence
of the scattering matrix.
To this end we consider the adiabatic problem 
when the parameters change periodically in time: 
$X_i(t) = X_i(t + 2\pi/\omega)$.
In this case we can expand the adiabatic scattering matrix 
into the Fourier series

\begin{equation}
\hat S_{ad}(E,t) = 
\sum_{n=-\infty}^{\infty} \hat S_{0,n}(E) e^{-in\omega t},
\label{Eq2_2}
\end{equation}

\noindent where

\begin{equation}
\begin{array}{l}
\hat S_{0,n}(E) = \frac{\omega}{2\pi} 
\int\limits_{0}^{2\pi/\omega} dt e^{in\omega t} \\
\ \\
\times\hat S_0(E,X_1(t),\dots,X_N(t)).
\end{array}
\label{Eq2_3}
\end{equation}

\noindent
The Fourier harmonics $S_{0,n}$ define the amplitudes of side bands 
for particles traversing the 
adiabatically oscillating scatterer with initial incident energy $E$.
Thus we can construct  
the adiabatic Floquet scattering matrix 
as follows

\begin{equation}
\hat S_{F,ad}(E_{n},E) =  
\hat S_{F,ad}(E,E_{-n}) =  
\hat S_{0,n}(E).
\label{Eq2_4}
\end{equation}

\noindent
The adiabatic 
Floquet scattering matrix consists thus for each incident 
energy of a block of dimension of $(2N_r n_{max})^{2}$ where 
$N_r$ is the number of leads and $n_{max}$ 
is the maximum number of side bands needed for 
an accurate description of the quantities of interest. 

The above equation allows us to formulate the following adiabaticity criterion:.
If the Floquet scattering matrix $\hat S_{F}$
changes only a little when the energy $E$ changes by 
$n_{max}\hbar\omega$
then the adiabatic approximation can be applied and 
$\hat S_{F}\approx\hat S_{F,ad}$.
Note that because of Eq.(\ref{Eq2_4})
the same criterion can be applied to the scattering matrix 
$\hat S_{0}(E)$.

Let us now calculate the directed charge 
current $I_{ad}$ and a heat flow $I_{E,ad}$
in the adiabatic limit.
Substituting Eq.(\ref{Eq2_4}) into  Eq.(\ref{Eq1_8})
we find the current flowing in lead $\alpha$ under 
an adiabatic change of parameters
(we put $f^{(in)} = f_{0}$):

\begin{equation}
\begin{array}{c}
 I_{ad,\alpha} = 
\frac{e\omega}{2\pi}\int\limits_{0}^{\infty} dE 
\left(-\frac{\partial f_0(E)}{\partial E} \right) \\
\ \\
\times
\sum\limits_{\beta}\sum\limits_{n}
n|S_{0,\alpha\beta,n}(E)|^2.
\end{array}
\label{Eq2_5a}
\end{equation}
We see that the current $I_{ad}$ is due to photon-assisted processes 
(the current depends on the intensity of side bands which 
is proportional to $|S_{0,\alpha\beta,n}(E)|^2$).
Therefore even in the "adiabatic" limit 
pumping is, strictly speaking, a nonadiabatic phenomenon.

Note that formally Eq. (\ref{Eq2_5a}) is obtained
for $\hbar\omega\ll k_BT$. 
However it gives the correct answer (in the adiabatic limit)
for an arbitrary ratio of the frequency $\omega$ to the temperature $T$.

Using Eq.(\ref{Eq2_2}) we can rewrite Eq.(\ref{Eq2_5a}) as follows

\begin{equation}
\begin{array}{c}
 I_{ad,\alpha} = 
i\frac{e\omega}{4\pi^2}\int\limits_{0}^{2\pi/\omega} dt
\int\limits_{0}^{\infty} dE 
\left(-\frac{\partial f_0(E)}{\partial E} \right) \\
\ \\
\times
\left( 
\frac{\partial\hat S_{0}(E,t)}{\partial t}
\hat S_{0}^{\dagger}(E,t)
\right)_{\alpha\alpha}.
\end{array}
\label{Eq2_5}
\end{equation}
This is Brouwer's formula \cite{Brouwer98}.
We emphasize that Eq.(\ref{Eq2_5a}) and Eq.(\ref{Eq2_5}) express
a (nonadiabatic in nature) pumped current in terms of the scattering 
matrix $\hat S_0$ for a "frozen" scatterer with time-dependent parameters
(see Eq.(\ref{Eq2_1})). 
This is correct if the scattering matrix $\hat S_0$
can be taken to be energy independent 
on the scale of the order of $n_{max}\hbar\omega$.
The last conclusion is correct
irrespective of the amplitude of the oscillating parameters. 
But $n_{max}$ might very well depend on whether one is in the 
weak amplitude limit or in the strong amplitude limit.

Next consider the heat flow $I_{E,ad,\alpha}$ produced by
the adiabatically oscillating scatterer in the lead $\alpha$.
Substituting the adiabatic scattering matrix Eq.(\ref{Eq2_4}).
into Eq.(\ref{Eq1_12}) we find 

\begin{equation}
\begin{array}{c}
 I_{E,ad,\alpha} = \frac{1}{h} \int\limits_{-\infty}^{\infty} d\epsilon 
\sum\limits_\beta \sum\limits_{n}
|S_{0,\alpha\beta,n}(\mu+\epsilon)|^2 \\
\ \\
\times 
\frac{
\left(\epsilon+\frac{n\hbar\omega}{2}\right)
\sinh\left(\frac{n\hbar\omega}{2k_BT} \right)
}{
\cosh\left(\frac{\epsilon}{k_BT} \right) +
\cosh\left(\frac{n\hbar\omega}{2k_BT} \right)}.
\end{array}
\label{Eq2_6a}
\end{equation}
At zero temperature (or if the temperature $T$ is less than the relevant 
energy scale for the scattering matrix $\hat S_{0}$) 
the above equation can be greatly simplified. 
In this case the heat flow can be expressed 
in terms of an on-shell, time-dependent adiabatic scattering matrix 
$\hat S_{0}(\mu,t)$:

\begin{equation}
\begin{array}{l}
I_{E,ad,\alpha} = -\frac{\hbar\omega}{8\pi^2}
\int\limits_{0}^{2\pi/\omega} dt \\
\ \\
\times\left( \frac{\partial^2\hat S_{0}(\mu,t)}{\partial t^2}
\hat S_{0}^{\dagger}(\mu,t)
\right)_{\alpha\alpha}.
\end{array}
\label{Eq2_7}
\end{equation}

\noindent 
This expression coincides with that 
given by Avron et al. \onlinecite{AEGS01}.

In the adiabatic limit it is possible 
to express the quantities of interest 
in terms of the stationary scattering matrix. 
This permits the very general and useful
expressions discussed in this paragraph. 
In contrast, in the non-adiabatic case, 
the information contained in the 
stationary scattering matrix is by definition 
not sufficient. As a consequence, in order to 
address the non-adiabatic case, 
we have now to consider 
a very specific example.

\section{Charge and heat flows  
produced by an oscillating double barrier}
\noindent
\label{ch}

To apply the general formalism of Sec.\ref{ga}
beyond the adiabatic approximation
we investigate a simple model \onlinecite{HN91}.
It is a one-dimensional scatterer consisting of
two 
delta function barriers oscillating with frequency $\omega$
and located at $x=-L/2$ and $x=L/2$.

To calculate the Floquet matrix we consider scattering
of electrons with energy $E$ coming from  $x = -\infty$.
The time-dependent Schr\"odinger equation for an electron
wave function $\Psi(x,t)$ reads
\begin{equation}
\begin{array}{l}
i\hbar\frac{\partial\Psi(x,t)}{\partial t} = 
\hat H(x,t)\Psi(x,t), \\
\ \\
\hat H(x,t) = - \frac{\hbar^2}{2m}
\frac{\partial^2}{\partial x^2} + V(x,t), \\
\  \\
V(x,t) = V_{1}(t)\delta(x+\frac{L}{2}) 
+ V_{2}(t)\delta(x-\frac{L}{2}), \\
\ \\
V_{i}(t) = V_{0i} + 2V_{1i}\cos(\omega t + \varphi_i), i = 1,2.
\end{array}
\label{Eq3_1}
\end{equation}
\noindent
Appendix A gives the exact solution of this model. 
We now use this solution to calculate 
the charge current flowing through
(as well as the heat current flowing from)
the oscillating double barrier potential 
(see Eq.(\ref{Eq3_1})) connecting two reservoirs
with the same temperature $T = 0$ and electrochemical potential $\mu$.
We use the units $2m = \hbar = e = 1$.
To be definite we consider the current $I_{1}$ ($I_{E,1}$)
flowing to the left ($x\to -\infty$).
The currents are calculated, using 
Eq.(\ref{Eq1_8}) and Eq.(\ref{Eq1_12})
with the Floquet scattering matrix determined as shown 
in Appendix A. 
We compare these currents with the adiabatic currents 
$I_{ad,1}$
and $I_{E,ad,1}$
using the Brouwer formula Eq.(\ref{Eq2_5})
and its analog Eq.(\ref{Eq2_7}).

For reference we write down the scattering matrix 
$\hat S_{0}(E)$ which we need to calculate the adiabatic currents:

\begin{equation}
 \hat S_{0} = \frac{e^{ikL}}{\Delta}
\left( 
\begin{array}{cc}
\xi + 2\frac{p_2}{k}\sin(kL) & 1 \\
\ \\
1 & \xi + 2\frac{p_1}{k}\sin(kL) 
\end{array}
\right).
\label{Eq4_1}
\end{equation}

\noindent
Here 
$k = \sqrt{\frac{2m}{\hbar^2}E}$;
$p_j = V_jm/\hbar^2$ (j = 1,2);
$\xi$ = $(1-\Delta)e^{-ikL}$;
$\Delta$ = $1 + \frac{p_1p_2}{k^2}(e^{2ikL} -1)$ + $i\frac{p_1+p_2}{k}$.
Of interest is a comparison of the exact Floquet 
scattering matrix and adiabatic theory in the limit of small 
and large frequencies. Furthermore, it is interesting to compare the 
symmetry conditions for pumping for the two theories.

\subsection{Adiabatic limit: $\omega\to 0$}
\indent
\label{al}

>From our calculations it follows that in the limit of
$\omega \to 0$ the adiabatic current $I_{ad}$ gives 
a good approximation for the pumped current
irrespective of the amplitude of the oscillating potentials $V_{1i}$.
However the criterion of adiabaticity depends strongly on 
the ratio $V_{1i}/\hbar\omega$.

As we have seen, the adiabatic approximation is valid
if the scattering matrix $\hat S_{0}$ 
is energy independent on the scale of the order of
$n_{max}\hbar\omega$.
Here $n_{max}$ (specified below) is the number of side bands
with noticeable amplitude excited by the oscillating scatterer.
Let us denote by $\delta E$ the energy scale over which 
the scattering matrix $\hat S_{0}$ changes significantly. 
For example, close to a resonance $\delta E$  
is of the order of the width $\delta$ of a resonance.
Then the adiabatic approximation is valid if
$n_{max}\hbar\omega \ll \delta E$. 
Hence the larger the number $n_{max}$ of excited side bands
the smaller the frequencies for which the adiabatic approximation is valid.

 \begin{figure}
  \vspace{3mm}
  \centerline{
   \epsfxsize8cm
   \epsffile{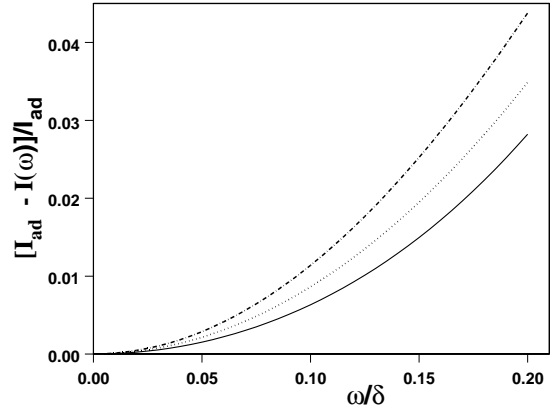}
             }
  \vspace{3mm}
  \nopagebreak
  \caption{Relative change of the pumped current
$[I_{ad} - I(\omega)]/I_{ad}$ 
as a function of the pump frequency
$\omega$  for three values of an oscillating potential
$V_{11}$ = $V_{12}$ = $0.02$ (solid line); $1$ (dotted line);
and $2$ (dash-dotted line)
close to the transmission resonance.
The frequency is measured in units of the width of the resonance
$\delta = 0.339$.
The parameters are: $L = 2\pi$; $\mu = 24.167$;
$V_{01} = V_{02} = 20$; $\varphi_1$ - $\varphi_2$ = $\pi/2$.
We use the units: $2m = \hbar = e = 1$.
}
\label{fig1}
\end{figure}

As it is well known \cite{Wagner}
the number $n_{max}$ of excited side bands
depends on 
the ratio of the amplitude of the oscillating potential $V_{1i}$
to the frequency $\omega$. 
In the small amplitude limit $V_{1i}\to 0$ only the first side bands
are excited: $n_{max} = 1$. 
But for strong pumping the number $n_{max}$ 
is large and the adiabatic approximation is valid only at smaller frequencies. 
As a consequence for a given finite frequency $\omega$
the deviation of the actual pumped current $I(\omega)$ 
from the adiabatic one $I_{ad}$ 
increases with increasing pumping amplitude.
This fact is illustrated in Fig.\ref{fig1}.

Before concluding this subsection we would like to emphasize the following.
For strong pumping the adiabatic approximation is still valid
at sufficiently small frequencies despite the excitation of 
a large number of side bands. The Fourier harmonics 
of the scattering matrix $\hat S_{0}$
define the amplitudes of side bands
for strong (Eq.(\ref{Eq2_4})) as well as for weak \cite{MB02}
{\it adiabatic} pumping.
Thus the scattering matrix $\hat S_{0}$ 
completely defines the kinetics and, in particular,
the quantum statistical correlation properties of an adiabatic pump.
The current noise of a weak (small amplitude) pump 
was considered in Ref.\onlinecite{MB02}.
The noise in a strong amplitude adiabatic pump will be presented elsewhere.

\subsection{Large frequency limit}
\indent
\label{lfl}

At large frequencies $\omega > \delta$ 
(where $\delta$ is the width of a resonance)
the pumped current $I$ 
differs considerably from the adiabatic one $I_{ad}$.

 \begin{figure}
  \vspace{3mm}
  \centerline{
   \epsfxsize8cm
   \epsffile{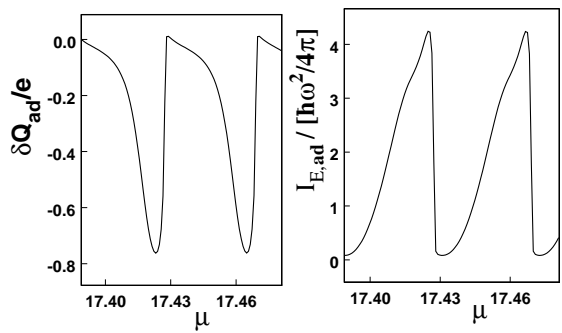}
             }
  \vspace{3mm}
  \nopagebreak
  \caption{
 The adiabatically pumped charge 
$\delta Q_{ad}$ = $2\pi I_{ad}/\omega$
(in units of an elementary charge $e$)
during a period (left panel)
and the adiabatic heat current $I_{E,ad}$ 
(in units of $\hbar\omega^2/(4\pi)$)
(right panel)
as a function of the Fermi energy $\mu$. 
The parameters are: 
$L = 200\pi$; 
$V_{01} = V_{02} = 20$; 
$V_{11} = V_{12} = 10$; 
$\varphi_1$ - $\varphi_2$ = $\pi/2$.
   }
\label{fig2}
\end{figure}

It is known 
that for a double barrier structure at $\omega\to 0$ 
the pumped dc current $I_{ad}$ 
\cite{WWG00,EWAL02}
and the generated heat flow $I_{E,ad}$ 
\cite{WW02}
show a resonance-like behavior as a function of the Fermi energy $\mu$
(see Fig.\ref{fig2}).
The pumped current and heat peak when the Fermi energy is close 
to the transmission resonance.
This is because in the adiabatic limit 
only the particles close to the Fermi level contribute to 
the charge (Eq.(\ref{Eq2_5})) and energy (Eq.(\ref{Eq2_7})) transfer.

With increasing  pumped frequency $\omega\gg\delta$ 
the particles within a wider energy interval come into play.
The dependence on the Fermi energy is smoothed away.
However a resonance-like dependence on the frequency $\omega$ arises. 

\begin{figure}
  \vspace{3mm}
  \centerline{
   \epsfxsize8cm
   \epsffile{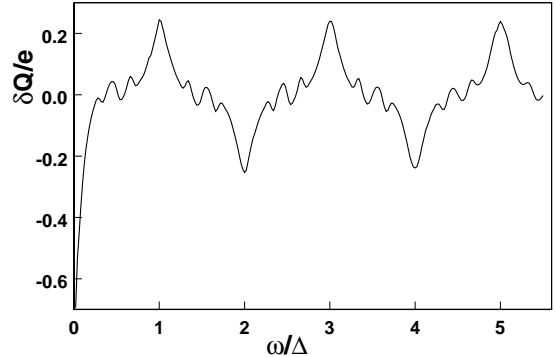}
             }
  \vspace{3mm}
  \nopagebreak
  \caption{
The charge $\delta Q$ = $2\pi I/\omega$ pumped during a cycle 
as a function of the frequency $\omega$ 
(in units of the separation $\Delta = 0.0417$
between the transmission resonances).
The Fermi energy is $\mu = 17.423$. 
The parameters are the same as in Fig.2.
   }
\label{fig3}
\end{figure}

In Fig.\ref{fig3} we depict the dependence of the charge 
$\delta Q$ = $(2\pi)I/\omega$
pumped during a cycle on the frequency $\omega$. 
The pumped charge peaks when the energy
quantum $\hbar\omega$ equals one (or several) level spacings $\Delta$ 
of a double barrier structure. 
In the example used for numerical calculations 
the level spacing $\Delta$ near the Fermi energy is constant
with good accuracy.
The smaller peaks correspond to many ($n = 2,3,\dots$) photon processes
for which $n\hbar\omega = m\Delta$ (m = 1,2,\dots). 
We see that the single photon processes dominate over 
the many photon processes.

At arbitrary ($n\hbar\omega \neq m\Delta$) but large
($\hbar\omega\gg\delta$)
frequencies, due to interference 
(inside the double barrier structure)
only the main component ($n=0$) 
of the Floquet state Eq.(\ref{Eq3_2}) 
has a significant amplitude between the barriers.
This component corresponds to the eigenfunction 
of the time independent problem
(with barriers $V_{01}$ and $V_{02}$).
The side bands ($n\neq 0$) do not participate 
in the transmission through the system
(more precisely, their contribution is small). 
As a result the pumped current is greatly reduced compared to the 
adiabatic case.
It should be noted that in the adiabatic case 
the side bands do contribute to the transmission
since for $\omega\ll \delta$ they lie at the same 
transmission resonance as the main component.

On the other hand 
at some particular values of a frequency $\hbar\omega = m\Delta$
the substates $\psi_{\pm k}$ ($k = 1, 2, 3, \dots$)
of the Floquet state Eq.(\ref{Eq3_2}) become large and 
additional channels for the transmission through the system open up. 
This leads to the increase of the pumped charge   
(see main peaks in Fig.\ref{fig3}).

Interestingly, as shown in Fig.\ref{fig3}), the pumped current reverses 
sign as a function of frequency. These sign reversals can be understood 
in the following way: 
If two potential barriers oscillate then the pump effect 
arises as a consequence of an interference between two amplitudes
\cite{WWG02}. 
The first amplitude ${\cal A}_1$
corresponds to particles which propagate through 
the double barrier structure and absorb (emit) the energy quantum
$\hbar\omega$ 
(or several quanta $n\hbar\omega$) 
at the first barrier.
The second amplitude ${\cal A}_2$ corresponds to the same propagation 
with absorption (emission) at the second barrier.
As it is known 
\cite{SMCG99,Brouwer98}
the adiabatically pumped current is an odd function of 
the phase difference 
of the oscillating potentials
$\Delta\varphi$ = $\varphi_1$ - $\varphi_2$ 
(see Eq.(\ref{Eq3_1})).
In fact this is the phase difference 
of the two amplitudes ${\cal A}_1$ and ${\cal A}_2$.
In the nonadiabatic case there exists an additional contribution
coming from the spatial phase difference
$$
\Delta\varphi_x \simeq  (k_1 - k_2)L.
$$
Thus at large frequencies the pumped current depends on
the sum $\Delta\varphi$ + $\Delta\varphi_x$.

To clarify the appearance of $\Delta\varphi_x$
let us consider particles with energy $E$ 
going from the left to the right and absorbing the quantum $\hbar\omega$.
The amplitude ${\cal A}_1$ corresponds to particles
which absorb an energy $\hbar\omega$ close to the left barrier ($V_1$)
and traverse the system at energy $E+\hbar\omega$. 
Hence $k_1\sim\sqrt{E+\hbar\omega}$.
On the other hand the amplitude ${\cal A}_2$ corresponds to particles
which absorb an energy $\hbar\omega$ close to the right barrier ($V_2$)
and thus traverse the system at energy $E$. They thus propagate with 
wave vector $k_2\sim\sqrt{E}$.
Because $k_1\neq k_2$ there is a phase difference
$\Delta\varphi_x$ between the amplitudes 
${\cal A}_1$ and ${\cal A}_2$.
Note that in the adiabatic case 
(i.e., at $\omega\to 0$)
$\Delta\varphi_x$ vanishes.

Now we show that in the case under consideration 
the phase difference $\Delta\varphi_x$ is at the origin of 
the sign reversal of the pumped current 
at consecutive peaks (see Fig.\ref{fig3}).
Close to the $m^{th}$ main peak of $I(\omega)$ the frequency 
is determined by 
$\hbar\omega\sim m\Delta$. 
Taking into account that 
only the particles with energy close 
to the resonance contribute to the transmission 
and mainly single photon processes are important 
we can estimate 
$\Delta\varphi_x$ as follows.
If the amplitude ${\cal A}_1$ corresponds to the propagation 
through some resonance (say $E_{l+m}$) then ${\cal A}_2$ 
corresponds to the propagation through the resonance $E_{l}$. 
At the $l^{th}$ transmission resonance it is 
$k^{(l)}L = l\pi$
the additional phase difference is $\Delta\varphi_x = m\pi$. 
Thus the sum $\Delta\varphi$ +  $\Delta\varphi_x$ 
changes by $\pi$ as we pass from one peak to another 
($\Delta m = 1$) thus giving rise to the sign reversal \cite{note2}.

 \begin{figure}
  \vspace{3mm}
  \centerline{
   \epsfxsize8cm
   \epsffile{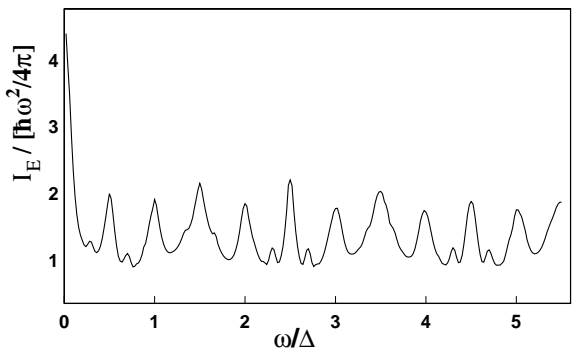}
             }
  \vspace{3mm}
  \nopagebreak
  \caption{
The heat current $I_{E}$ (in units of $\hbar\omega^2/(4\pi)$)
as a function of the frequency $\omega$.
The parameters are the same as in Fig.3.
   }
\label{fig4}
\end{figure}

In addition interference inside 
the double barrier structure manifests itself 
as a resonance-like dependence of the heat production (see Fig.\ref{fig4}).
We can see that $I_{E}(\omega)$ contains
additional peaks at $\hbar\omega$ = $(m+\frac{1}{2})\Delta$.
This fact emphasizes the striking difference between 
the processes which are responsible for charge pumping  
and for heat production \cite{MB02}.

\subsection{Quantum pumping and broken symmetry}
\indent
\label{bs}

Our results show that 
the symmetry of the scattering problem 
is important for the pump effect.

The quantum pump effect arises due to the different, interfering 
excitation histories of carriers traversing the 
sample \cite{rectification}. 
To "extract" directed currents from the uniform environment
the left-right symmetry (LRS) for carriers traversing the 
sample needs to be broken. 
There are, at least, two ways to break the LRS.
The first one is breaking the spatial symmetry (SS).
The second one is breaking the time reversal symmetry (TRS).

We restrict our considerations to systems which in the absence 
of the time-dependent perturbations needed for pumping 
are in an equilibrium state. 
In particular this means that without the presence of magnetic fields (etc.)
the system is time reversal invariant. 
Note that the time independent scattering problem is insensitive
to the presence (absence) of the spatial symmetry of the scatterer
(in the sense that the transmission probability is invariant
under the spatial inversion $x\to -x$ 
irrespectively of the SS of the scattering potential).

Interestingly, the symmetry conditions 
for pumping depend on the frequency.
In the adiabatic limit $\omega\to 0$ (see Sec.\ref{aa})
the scattering problem is fully characterized by the 
scattering matrix $\hat S_{0}$. In contrast to the conductance, 
adiabatic pumping is sensitive to the symmetry of 
the scattering matrix. To have an adiabatic pumping 
effect we need to break both the spatial symmetry and 
the time-reversal symmetry. 

It is important to distinguish the symmetry of the 
equilibrium problem and the symmetry of the full 
problem. In the adiabatic case, the symmetry of the 
time-independent problem is irrelevant: What matters, 
is that the system in presence of perturbations 
breaks both the spatial symmetry and the time
reversal symmetry. In many examples (for instance the two
barrier problem considered above) it is not possible 
to break the time-reversal symmetry without at the 
same time breaking the spatial symmetry. However, 
examples which are spatially symmetric 
and have broken time-reversal invariance can be constructed. 
For instance, we could consider the problem of two barriers 
oscillating in synchronism and as a second perturbation 
which oscillates with a phase lag choose the potential 
between the two barriers. 
This would be an example of a two parameter problem 
which does not generate a pumped current. 
The important point we would like to emphasize 
is that the symmetry conditions for the non-adiabatic pump 
are different. 

In the strongly nonadiabatic limit the scattering problem is described 
by the Floquet scattering matrix $\hat S_{F}(E^{(out)}, E^{(in)})$
dependent on both the incident $E^{(in)}$ 
and outgoing $E^{(out)}$ energies.
In general, the Floquet scattering matrix is sensitive to the
spatial symmetry of the scatterer. 
If the scattering problem is without SS 
(i.e., when the scattering potential is not invariant under 
the change $x\to -x$) then 
$$
\begin{array}{c}
|S_{F,\alpha\beta}(E^{(out)}, E^{(in)})|^2 \\
\ \\
\neq 
|S_{F,\beta\alpha}(E^{(out)}, E^{(in)})|^2,
\end{array}
$$
and the oscillating scatterer can pump a dc current
(see Eq.(\ref{Eq1_9})).
Thus to obtain a pump effect in the strongly nonadiabatic limit
there are, at least, two possibilities.
Either the spatial symmetry or the time reversal symmetry
(or both SS and TRS together) have to be broken.
The role of spatial symmetry breaking
for the nonadiabatic pump effect 
was emphasized in Ref.\onlinecite{WWG02}.

\begin{figure}
  \vspace{3mm}
  \centerline{
   \epsfxsize8cm
   \epsffile{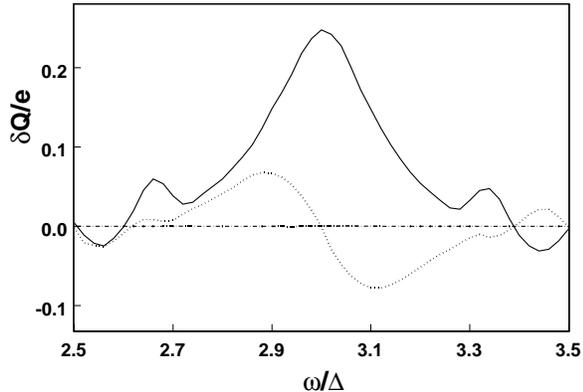}
             }
  \vspace{3mm}
  \nopagebreak
  \caption{
The charge $\delta Q$ pumped during a cycle 
as a function of the frequency $\omega$.
Three cases are presented:
(i) two potentials $V_{11} = V_{12} = 10$ oscillate out of phase
$\varphi_1$ - $\varphi_2$ = $\pi/2$ (solid line);
(ii) only single potential oscillates: $V_{11} = 10$, $V_{12} = 0$
(dotted line);
(iii) two potentials $V_{11} = V_{12} = 10$ oscillate in phase
$\varphi_1$ = $\varphi_2$ (dash-dotted line).
The parameters are the same as in Fig.3.
   }
\label{fig5}
\end{figure}

The above reasoning is illustrated in Fig.\ref{fig5}
where the dependence of a pumped current on the frequency is shown 
for three generic situations.
First, two potentials oscillate with the same amplitude but out of phase 
(solid line). 
In this case the perturbation breaks both the spatial 
symmetry and the time reversal symmetry. 
Second, only one potential oscillates (dotted line). 
The TRS is present but the SS is dynamically broken. 
The system is able to pump a dc current like in the previous case.
Third, two potentials oscillate in phase and with the same amplitudes
(dash-dotted line).
Now both the TRS and SS are preserved and 
the system under consideration does not exhibit a directed current.

\section{Discussion and conclusion}
\indent
\label{dc}

In this work we have developed the Floquet scattering matrix approach 
to parametric pumping 
in phase coherent mesoscopic systems of noninteracting electrons.
We have calculated the distribution function, the dc current and 
the heat flow at arbitrary pumping amplitudes and frequency.

The Floquet scattering matrix describes naturally
the existence of the side bands 
$E_n = E +n\hbar\omega$
of particles leaving the pump. 
These side bands 
correspond to nonequilibrium particles generated by the pump
which carry the heat from the oscillating scatterer 
to the reservoirs and (under some conditions) transfer charge
between the reservoirs \cite{MB02}.
If the pumping amplitude is small only the first 
($n = \pm1$) side bands are excited.
But for strong pumping the number of excited side bands
is large: $n_{max}\gg 1$. 
This number together with the frequency defines the energy 
scale $n_{max}\hbar\omega$
characteristic of the pump problem.
In particular this is important for the analysis of the conditions 
under which the adiabatic approach to pumping is valid.
If the scattering matrix is energy independent on the scale of the
order of $n_{max}\hbar\omega$ 
then the adiabatic approximation can be applied.
In this case the elements of the Floquet scattering matrix
are given by
the corresponding Fourier components 
of the stationary scattering matrix $\hat S_{0}$ (see Eq.(\ref{Eq2_4}))
with parameters taken to be dependent on time.

The existence of the pump effect is 
directly related to 
the symmetry of the scattering problem.
In the adiabatic case ($\omega\to 0$) 
only a scatterer without spatial and time reversal symmetry 
can produce a directed current. 
This conclusion applies irrespectively of the 
(stationary) spatial symmetry of the scatterer.
On the other hand in the nonadiabatic case 
(at large pumping frequency)
to achieve pumping we need a scatterer with 
either broken spatial or time reversal symmetry.
Hence at large operating frequencies 
even the scatterer with a single oscillating parameter
can show a pump effect if only the spatial symmetry is broken 
\cite{WWG02}.

To emphasize the main physics underlying pumping 
we have considered an exactly solvable model.
Namely a one-dimensional scatterer consisting of
two oscillating delta-function barriers Eq.(\ref{Eq3_1})
separated by a distance $L$.

If two barriers oscillate out of phase
(broken TRS)
the basic process leading to the pump effect is 
an interference of two quantum mechanical amplitudes. 
These amplitudes ${\cal A}_1$ and ${\cal A}_2$ corresponds to particles 
which traverse the scatterer in the same direction
but gain (or lose) a modulation energy 
in the vicinity either of the first 
or of the second barrier, respectively \cite{WWG02}. 
This becomes clearer at large frequencies $\omega$
because of the following.
When the energy quantum $\hbar\omega$ equals one 
(or several, say $m$) level spacing $\Delta$ of a double barrier structure
the pumped current peaks (see Fig.\ref{fig3}). 
The particles 
with energy close to the resonance of a double
barrier structure give the main contribution to the pump current.
Thus the amplitudes ${\cal A}_1$ and ${\cal A}_2$ correspond to 
traversal of the system through different 
(say, $(l+m)^{th}$ and $l^{th}$) resonances.
This gives rise to an additional phase difference 
$\Delta\varphi_x$ = $(k_1-k_2)L$
between the amplitudes ${\cal A}_1$ and ${\cal A}_2$.
For the resonant double barrier considered here 
the phase difference 
between consecutive resonances is $\pi$ and 
thus $\Delta\varphi_x = m\pi$. 
As a result the consecutive peaks of the pumped current
have opposite sign (see Fig.\ref{fig3}).

We conclude that if the time reversal symmetry is broken
(i.e., $\Delta\varphi\neq 0$),
the nonadiabatic pumped current
contains direct information on the phase of the transmission
coefficient through the mesoscopic sample.
We emphasize that here we have a single-connected geometry 
in contrast with the case where the mesoscopic sample (a quantum dot)
is embedded in one arm of the ring \cite{YHMS}. 

We have also considered the effect of a spatial asymmetry.
If only one potential (say, $V_1$) oscillates 
then the spatial symmetry is (dynamically) broken.
In this case the (nonadiabatic) pump effect 
arises because the probability for exciting side bands 
$|S_{F,\alpha\beta}(E_{n},E)|^2$
depends on the direction 
(i.e., on the order of indexes $\alpha$ and $\beta$) 
(see Eq.(\ref{Eq1_9})). 
The particles going from the left to the right
first absorb (emit) the energy 
$n\hbar\omega$ = $E_n - E$ 
and then pass through the double barrier structure (at energy $E_n$).
Thus the corresponding probability is 
$$
|S_{F,21}(E_{n},E)|^2 \sim T(E_n).
$$
Here $T(E)$ is a transmission probability.
On the other hand 
the particles going from the right to the left 
pass the system before they absorb (or emit) the energy
in the vicinity of the oscillating barrier $1$.
Hence we have 
$$|S_{F,12}(E_{n},E)|^2 \sim T(E).
$$
Thus the pumped current Eq.(\ref{Eq1_9}) depends on the differences
$T(E) - T(E\pm n\hbar\omega)$. 

Note that for a system with equidistant spectrum 
(that is the case under consideration) the transmission probability
$T(E)$ is periodic in energy $E$ with the period of $\Delta$.
Hence if
$\hbar\omega = m\Delta$ 
then it is $T(E)\approx T(E\pm\hbar\omega)$ and the current is 
close to zero in accordance with Fig.\ref{fig5} 
(a dotted curve; $m = 3$).

We have presented a scattering theory of quantum pumping 
based on the Floquet theorem.
This approach allows a description of the kinetics
of both adiabatic and nonadiabatic quantum pumps.
In particular the quantum statistical correlation properties (noise) 
can be considered in analogy with the weak amplitude adiabatic quantum pump
\cite{MB02}. The investigation of 
noise is especially important in the context of quantized
(i.e., noiseless) charge pumping 
\cite{AEGS01,Alekseev,MB02}
and will be presented elsewhere.

\ \\
\centerline{\bf ACKNOWLEDGEMENT}
\indent

This work is supported by the Swiss National Science Foundation.

\section{Appendix A: Oscillating double barrier. An exact solution}
\indent
\label{es}

In this Appendix we determine the Floquet scattering 
matrix of the Schr\"odinger equation (\ref{Eq3_1})
of two oscillating delta function potentials 
located at $x=-L/2$ and $x=L/2$.
To find the Floquet matrix we consider scattering
of electrons with energy $E$ coming from  $x = -\infty$.
To solve (\ref{Eq3_1}) we use the Floquet functions method 
\cite{single}. Since the Hamiltonian $\hat H$ depends on
time, the system has no stationary eigenstates.
However, the Floquet theorem tells us that 
because the Hamiltonian is periodic in time
the eigenstates  of Eq.(\ref{Eq3_1})
can be represented as a superposition of 
wave functions with energies shifted by $n\hbar\omega$:
\begin{equation}
\Psi_{E}(x,t) = e^{-iEt/\hbar}\sum^\infty_{n=-\infty}\psi_n(x)
e^{-in\omega t}.
\label{Eq3_2}
\end{equation}
\noindent
Away from the points $x=-L/2$ and $x=L/2$ the functions 
$\psi_n(x)$ are a superposition of plane waves
\begin{equation}
\psi_n(x) = a_n e^{ik_n x} + b_n e^{-ik_n x},
\label{Eq3_3}
\end{equation}
\noindent
where 
$$
k_n = \sqrt{\frac{2m}{\hbar^2}\left(E + n\hbar\omega \right)}
$$
with $Re[k_n]\geq  0$ and $Im[k_n]\geq 0$.
At $x = -L/2$ and $x = L/2$ we have the boundary conditions
\begin{equation}
\begin{array}{c}
\Psi_{E}(-\frac{L}{2}-0,t) = \Psi_{E}(-\frac{L}{2}+0,t), \\
\ \\
\Psi_{E}(\frac{L}{2}-0,t) = \Psi_{E}(\frac{L}{2}+0,t), \\
\ \\
{\left.\frac{\partial\Psi_E(x,t)}{\partial x}\right|_{x=-\frac{L}{2}+0}} -
{\left.\frac{\partial\Psi_E(x,t)}{\partial x}\right|_{x=-\frac{L}{2}-0}} \\
\ \\
 = \frac{2m}{\hbar^2}V_{01}(t)\Psi_{E}(-\frac{L}{2},t),  \\
\ \\
{\left.\frac{\partial\Psi_E(x,t)}{\partial x}\right|_{x=\frac{L}{2}+0}} -
{\left.\frac{\partial\Psi_E(x,t)}{\partial x}\right|_{x=\frac{L}{2}-0}} \\
\ \\
 = \frac{2m}{\hbar^2}V_{02}(t)\Psi_{E}(\frac{L}{2},t).  \\
\end{array}
\label{Eq3_4}
\end{equation}
In the case under consideration 
(i.e., when an electron with energy $E$ comes from $x=-\infty$) 
the functions $\psi_n(x)$ are 
\begin{equation}
\begin{array}{l}
\psi_n(x<0) = \delta_{n,0} e^{ik_n x} + R_n e^{-ik_n x}, \\
\ \\
\psi_n(0 < x < L) = A_n e^{ik_n x} + B_n e^{-ik_n x}, \\
\ \\
\psi_n(x > L) = T_n e^{ik_n x}.
\end{array}
\label{Eq3_5}
\end{equation}
\noindent 
Here the coefficients $R_n$ and $T_n$ for propagating modes
(for which $E_n > 0$) 
are the amplitudes of reflection from or
transmission through the double barrier system 
absorbing ($n > 0$) or emitting
($n < 0$) an energy $|n|\hbar\omega$, respectively.

Substituting Eq.(\ref{Eq3_2}) and Eq.(\ref{Eq3_5}) into the boundary
conditions Eq.(\ref{Eq3_4}) we obtain
\begin{equation}
\begin{array}{l}
R_n = \left(A_n - \delta_{n,0} \right) e^{-ik_n L} + B_n, \\
\ \\
T_n = A_n + B_n e^{-ik_n L},
\end{array}
\label{Eq3_6}
\end{equation}
\noindent 
where the coefficients $A_n$ and $B_n$ are subject to 
recursive equations. 
It is convenient to represent these equations in a matrix form
\begin{equation}
\begin{array}{c}
 \hat U_{n} 
\left(
\begin{array}{l}
A_n \\
B_n
\end{array}
\right) 
+ \delta_{n,0}
\left(
\begin{array}{c}
-ik_0e^{-ik_0\frac{L}{2}} \\
0
\end{array}
\right) 
\\
\ \\
 = \hat D_{n+1} 
\left(
\begin{array}{l}
A_{n+1} \\ 
B_{n+1}
\end{array}
\right) 
+ \hat I_{n-1} 
\left(
\begin{array}{l}
A_{n-1} \\
B_{n-1}
\end{array}
\right).
\end{array}
\label{Eq3_7}
\end{equation}
\noindent 
Here we have introduced the matrices
\begin{equation}
\hat U_n = 
\left(
\begin{array}{lr}
(ik_n - p_{01})e^{-ik_n\frac{L}{2}} & -p_{01}e^{ik_n\frac{L}{2}} \\
\ \\
-p_{02}e^{ik_n\frac{L}{2}} & (ik_n - p_{02})e^{-ik_n\frac{L}{2}} \\
\end{array}
\right),
\label{Eq3_8}
\end{equation}
\begin{equation}
\hat D_{n+1} = 
\left(
\begin{array}{lr}
 p_{11}e^{i\varphi_1}e^{-ik_{n+1}\frac{L}{2}} & 
 p_{11}e^{i\varphi_1}e^{ik_{n+1}\frac{L}{2}} \\
\ \\
 p_{12}e^{i\varphi_2}e^{ik_{n+1}\frac{L}{2}} & 
 p_{12}e^{i\varphi_2}e^{-ik_{n+1}\frac{L}{2}} \\
\end{array}
\right),
\label{Eq3_9}
\end{equation}
\begin{equation}
\hat I_{n-1} = 
\left(
\begin{array}{lr}
 p_{11}e^{-i\varphi_1}e^{-ik_{n-1}\frac{L}{2}} & 
 p_{11}e^{-i\varphi_1}e^{ik_{n-1}\frac{L}{2}} \\
\ \\
 p_{12}e^{-i\varphi_2}e^{ik_{n-1}\frac{L}{2}} & 
 p_{12}e^{-i\varphi_2}e^{-ik_{n-1}\frac{L}{2}} \\
\end{array}
\right),
\label{Eq3_10}
\end{equation}
\noindent 
where the parameters are:
$p_{ij} = V_{ij}m/\hbar^2$, $i=0,1$, $j=1,2$.

To solve the equation (\ref{Eq3_7}) we  
have generalized the method used in Ref.\onlinecite{MR01}
for a single oscillating delta-function potential.

First of all we consider Eq.(\ref{Eq3_7}) at $n > 0$.
We suppose that there exist matrices $\hat X_{n}$ such that
\begin{equation}
\left(
\begin{array}{l}
A_{n} \\ 
B_{n}
\end{array}
\right) 
= \hat X_{n}
\left(
\begin{array}{l}
A_{n-1} \\ 
B_{n-1}
\end{array}
\right).
\label{Eq3_11}
\end{equation}
\noindent 
Then substituting the above constraint into Eq.(\ref{Eq3_7})
we obtain a simple recursive equation for the matrices $\hat X_{n}$
($n > 0$)
\begin{equation}
\hat X_{n} = \left(
\hat U_{n} - \hat D_{n+1} \hat X_{n+1}
\right)^{-1} \hat I_{n-1}.
\label{Eq3_12}
\end{equation}
\noindent
Using these matrices we can express all the coefficients
$A_{n}$, $B_{n}$ ($n > 0$) in terms of $A_0$ and $B_0$ only
\begin{equation}
\left(
\begin{array}{l}
A_{n} \\ 
B_{n}
\end{array}
\right)
= \hat X_{n}\dots\hat X_1
\left(
\begin{array}{l}
A_{0} \\ 
B_{0}
\end{array}
\right),~~~ n > 0.
\label{Eq3_13}
\end{equation}

Further we consider $n < 0$ and introduce the matrices $\hat Y_{n}$
\begin{equation}
\left(
\begin{array}{l}
A_{n} \\ 
B_{n}
\end{array}
\right) 
= \hat Y_{n}
\left(
\begin{array}{l}
A_{n+1} \\ 
B_{n+1}
\end{array}
\right).
\label{Eq3_14}
\end{equation}
\noindent 
The corresponding recursive equation for the matrices $\hat Y_{n}$
is 
\begin{equation}
\hat Y_{n} = \left(
\hat U_{n} - \hat I_{n-1} \hat Y_{n-1}
\right)^{-1} \hat D_{n+1}.
\label{Eq3_15}
\end{equation}
\noindent
Using these matrices we express the coefficients
$A_{n}$, $B_{n}$ ($n < 0$) in terms of $A_0$ and $B_0$:
\begin{equation}
\left(
\begin{array}{l}
A_{n} \\ 
B_{n}
\end{array}
\right)
= \hat Y_{n}\dots\hat Y_{-1}
\left(
\begin{array}{l}
A_{0} \\ 
B_{0}
\end{array}
\right),~~~ n < 0.
\label{Eq3_16}
\end{equation}
As a final step we consider Eq.(\ref{Eq3_7}) at $n = 0$.
After simple manipulations we find the coefficients
$A_0$ and $B_0$:
\begin{equation}
\begin{array}{c}
\left(
\begin{array}{l}
A_{0} \\ 
B_{0}
\end{array}
\right)
= 
\left(
\hat U_{0} - \hat D_{1}\hat X_{1} - \hat I_{-1}\hat Y_{-1}
\right)^{-1} \\
\ \\
\times
\left(
\begin{array}{c}
ik_0e^{-ik_0\frac{L}{2}} \\ 
0
\end{array}
\right).
\end{array}
\label{Eq3_17}
\end{equation}
Thus, using the solutions of the recurrent equations
Eq.(\ref{Eq3_12}) and Eq.(\ref{Eq3_15}) we can calculate 
all the coefficients $A_{n}$, $B_{n}$ 
(see Eq.(\ref{Eq3_13}), Eq.(\ref{Eq3_16}), and Eq.(\ref{Eq3_17})).
Note that in each particular case 
we need to take into account only
the limited number $|n| < n_{max}$ of side bands
and thus we can put 
$\hat X_{n_{max}+1}\approx 0$ and 
$\hat Y_{-n_{max}-1}\approx 0$.

The coefficients $R_{n}$ and $T_{n}$ 
can be calculated with the help of Eq.(\ref{Eq3_6}). 
For the propagating modes 
($E_n\equiv E + n\hbar\omega > 0$) of interest here,
these coefficients 
determine the elements of the Floquet scattering matrix 
\begin{equation}
\begin{array}{c}
 |S_{F,11}(E_n,E)|^{2} = \frac{k_n}{k_0} |R_n|^{2}, \\
\ \\
 |S_{F,21}(E_n,E)|^{2} = \frac{k_n}{k_0} |T_n|^{2}. 
\end{array}
\label{Eq3_18}
\end{equation}
\noindent Here the indexes $1$ and $2$ 
correspond to the left and right reservoirs, respectively. 
To obtain the matrix elements
$S_{F,22}$ and $S_{F,12}$ we need to solve the same problem 
with plane waves coming from the right.

\end{multicols}

\begin{thebibliography}{11}
\bibitem{note1} 
The crossover from adiabatic to non-adiabatic transmission 
through a tunnel barrier with an additional small amplitude oscillating 
potential was used in Ref. \onlinecite{BL82} to find the traversal time
for tunneling. This yields a time-scale that differs form 
the more widely used Wigner and Wigner-Smith times. 
We refer the interested reader to \onlinecite{muga}.
\bibitem{single}
The same problem but with a single oscillating delta function barrier
was considered in Ref.\onlinecite{MR01}.
\end{thebibliography}
\end{document}